\documentclass[a4paper,11pt]{article}
\usepackage{pos}
\usepackage{blindtext}
\usepackage{url}
\usepackage{mathtools}
\usepackage{bbm}
\usepackage{xcolor}
\usepackage{bbold}
\usepackage{caption}
\usepackage{subcaption}
\usepackage{wrapfig}
\usepackage{fancyhdr}
\usepackage{physics}
\usepackage{slashed}
\usepackage{appendix}
\usepackage{pifont}
\usepackage{float}
\usepackage{afterpage}
\usepackage{bm}
\usepackage{sidecap}

\title{Dissociation and regeneration of charmonia within microscopic Langevin simulations}

\author*[a]{Naomi Oei}
\author[b]{Juan M. Torres-Rincon}
\author[a]{Hendrik van Hees}
\author[a]{Carsten Greiner}

\affiliation[a]{Institut für Theoretische Physik, Johann Wolfgang Goethe-Universität, Max-von-Laue-Strasse 1, 60438 Frankfurt am Main, Germany}
\affiliation[b]{Departament de Física Quàntica i Astrofísica and Institut de Ciències del Cosmos (ICCUB), Facultat de Física, Universitat de Barcelona, Barcelona, Spain}

\emailAdd{oei@itp.uni-frankfurt.de}

\abstract{
We present a classical model to study the formation of charmonia, as well as dissociation and regeneration processes of heavy-quark bound states in the quark gluon plasma using Langevin simulations. The charm and anticharm quarks are described as Brownian particles in the background medium of light quarks and gluons and interact among them over a Coulomb-like screened potential to form bound states, which can dissociate again due to interactions with the medium. Box simulations at fixed temperature and volume are used to verify that the system reaches the expected thermal distribution in the equilibrium limit and to test bound state properties. The medium evolution is then parametrized by a boost-invariant fireball. In this configuration, the elliptic flow of charm and anticharm quarks as well as of charmonia is studied at RHIC and LHC energies.
}

\FullConference{%
  QNP2024 - The 10th International Conference on Quarks and Nuclear Physics, \\
   8-12 July, 2024 \\
   Barcelona, Spain 
}

\begin{document}

\maketitle

\section{Introduction}

The study of heavy quarks (HQ) and quarkonium states serves as a crucial tool for understanding the medium formed in relativistic heavy-ion collisions (HICs). Matsui and Satz first proposed in 1986 that the formation of $J/\psi$ and other bound states is suppressed in the quark-gluon plasma (QGP) due to screening effects of color charges~\cite{MatsuiSatz}. In this hot and dense medium, the interaction necessary to bind a charm and anticharm quark is weakened by the presence of light quarks and gluons. This suppression was confirmed experimentally at the SPS and later at RHIC, see, e.g.,~\cite{NA50:1997hlx,PHENIX:2006gsi}.

Heavy quark-antiquark pairs are generated in the early stages of  relativistic HICs through hard scatterings, and their number remains largely conserved during the medium evolution. As they interact with the QGP, quarkonium states can dissociate through elastic scatterings with medium particles. At sufficiently high energies, these dissociation processes are countered by the process of regeneration, where new bound states are formed from HQ recombination. However, the statistical hadronization model (SHM) has successfully predicted quarkonium yields at the hadronization phase, where the hadron production at the freeze-out stage is described within a statistical mechanics framework. The process is treated as instantaneous, determined by the chemical freeze-out temperature and baryochemical potential, assuming the source volume has reached statistical equilibrium and can be modeled using a grand-canonical ensemble.

In this work, we develop a classical description of HQ dynamics using relativistic Langevin simulations. Heavy quarks are treated as Brownian particles, whose random motion results from scatterings with medium constituents. The interaction between HQs is modeled through a complex potential: the real part represents a screened Coulomb attraction allowing for bound-state formation, while the imaginary part accounts for medium-induced dissociation effects. This framework captures the dissociation and regeneration of quarkonium throughout the QGP evolution.
 
In Sec.~\ref{sec:formalism} we describe the formalism of the HQ propagation using the Langevin equation as well as the interaction of HQ pairs to form charmonia through a potential. In Sec.~\ref{sec:results} we present our results, both for simulations in a cubic box at constant temperature and for an expanding system using a dynamical fireball. In Sec.~\ref{sec:outlook} we summarize our results and provide a short outlook.
 
\section{Formalism}
\label{sec:formalism}

The HQs are treated as Brownian particles propagating through the medium, with their random movement driven by collisions with light background particles. This approach can be modeled with a Fokker-Planck equation, with transport coefficients (drag forces and diffusion coefficients) encoding the forces acting on the HQs. We consider the static limit of this equation, $\bm{p} \xrightarrow[]{} 0$, in which a unique independent drag coefficient, $\gamma$, is needed. Other transport coefficients are connected via the relativistic dissipation-fluctuation relation. Details are contained in Ref.~\cite{Oei:2024zyx}.

The Fokker-Planck equation is microscopically realized with Langevin equation simulations,
\begin{equation}
\left\{ 
\begin{array}{ccc}
    d \bm{x} & =& \bm{p}/E(p) \ dt \\
    d \bm{p} & = &-\gamma \bm{p} dt + \bm{F} (|\bm{x} - \bm{\bar{x}}|;T) dt  + \sqrt{2\gamma E(p) T dt} \ \bm{\rho}\ ,
\end{array}
\right.
\end{equation}
where $\bm{F} (|\bm{x} - \bm{\bar{x}}|,T)$ defines the force between a HQ pair resulting from the HQ potential ($\bm{x}$ and $\bm{\bar{x}}$ are the positions of the heavy quark and antiquark, respectively), $T$ is the temperature of the medium, $\gamma$ denotes the drag coefficient and $\bm{\rho}$ is a Gaussian-distributed random variable. 

The HQ potential is adopted from the formalism developed in Ref.~\cite{Blaizot} and given as a Coulomb-like complex potential, 
\begin{equation}
    \mathcal{V}(r) = - \frac{g^2}{4\pi} m_D - \frac{g^2}{4\pi}  \frac{\exp(-m_D r)}{r} - i \frac{g^2T}{4 \pi}\phi(m_D r) \ , \label{eq:Blaizot}
\end{equation}
where $r=|\bm{x} - \bm{\bar{x}}|$ is the relative distance between the heavy quark and antiquark. The first term represents a constant self-energy contribution. The second Coulomb-like term describes the attractive interaction between a HQ pair, screened by the Debye mass $m_D$ of the light constituents of the medium. Meanwhile, the imaginary part accounts for the momentum loss during scatterings with medium particles. The explicit form of the function $\phi(m_Dr)$ is given in~\cite{Blaizot,Oei:2024zyx}.

The potential of Eq.~(\ref{eq:Blaizot}) allows for dissociation processes, driven both by the screening of the interaction and by collisions with plasma particles.  The strong coupling is parametrized by
\begin{equation}
    g^2 (T) = 4 \pi \alpha_s(T) = 4\pi \alpha_s(T_c) / \left[1+ C \ln \left( T/T_c \right) \right] \ , 
\end{equation}
with $C=0.76$ and $T_c = 160\ \text{MeV}$. 

In this work, we choose $\alpha_s = 0.7$~\cite{Oei:2024zyx}, as will be motivated in Sect.~\ref{sec:results}. The drag coefficient $\gamma$ results from the second derivative of the imaginary part of the potential and is given by
\begin{align}
    \gamma & = \frac{m_D^2g^2}{24\pi M_c} \left[ \ln \left( 1+ \frac{\Lambda^2}{m_D^2} \right) - \frac{\Lambda^2}{ \Lambda^2+m_D^2} \right] \ ,  \label{eq:drag}
\end{align}
where we introduce a cut-off $\Lambda = 4 \ \text{GeV}$, as proposed in Ref.~\cite{Blaizot}. Plots of the real part of the potential (\ref{eq:Blaizot}) as a function of the HQ distance as well as the drag coefficient (\ref{eq:drag}) as a function of the temperature are displayed in Ref.~\cite{Oei:2024zyx}.

\section{Results}
\label{sec:results}

The behavior of the HQs can be evaluated through box simulations at fixed volume and temperature~\cite{Oei:2024zyx}. As a first step, we aim to examine the formation of bound states. For this purpose, we adopt the criterion that a charm-anticharm pair is considered to be bound if its relative energy is smaller than zero. This is determined by subtracting the center-of-mass energy from the total energy of the system, $E_{c \bar{c}} = E_c + E_{\bar{c}}  + V(r) - E_{\textrm{tot}} $, where $V(r)=\textrm{Re } \mathcal{V}(r)$. This condition is checked at every time step, and for every pair combination.

\begin{figure}[H]
    \centering
    \includegraphics[width=0.45\linewidth]{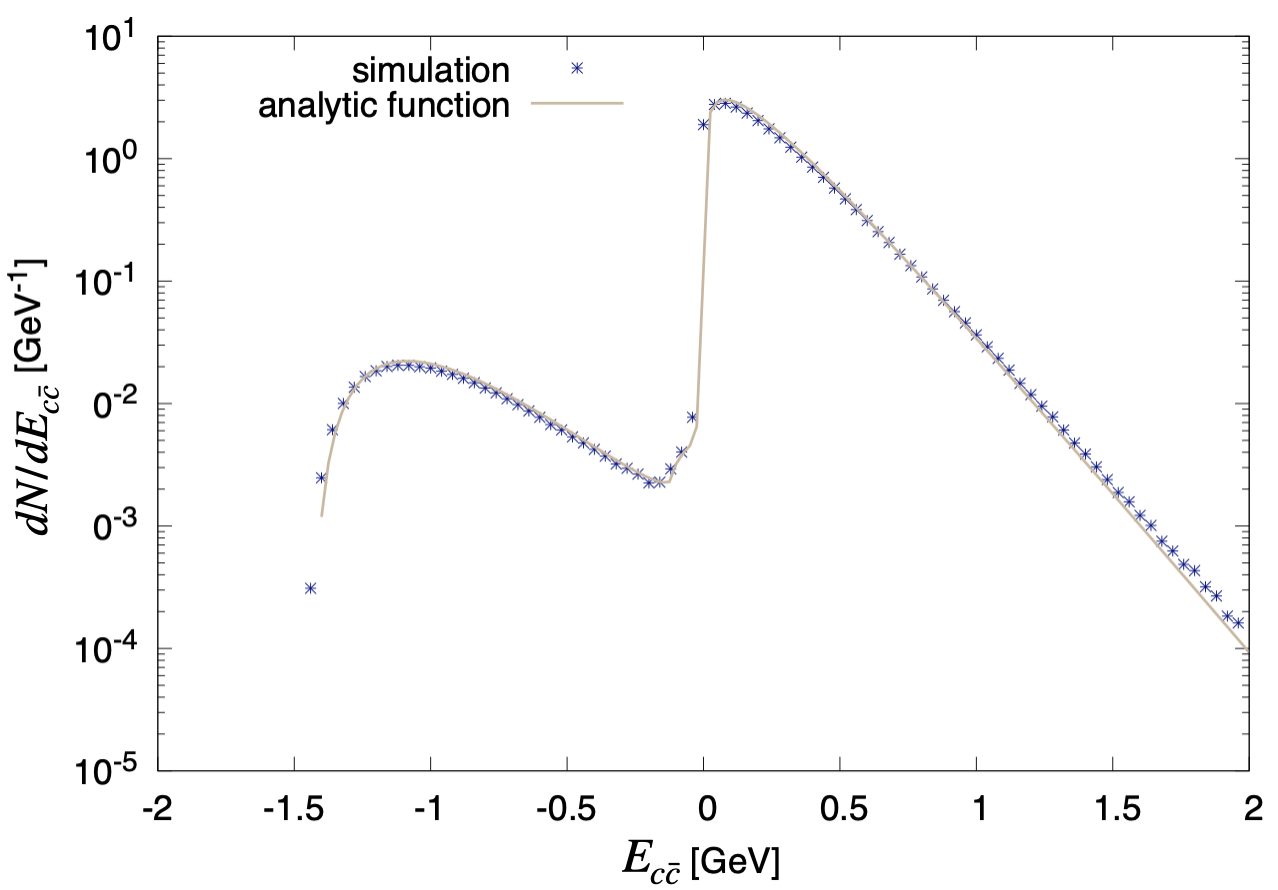}
    \caption{Simulation of the distribution of the relative energy of the heavy-quark pair in a cubic box with volume $V=(8\ \text{fm})^3$ and temperature $T=160\ \text{MeV}$, compared to the analytic expectation, Eq.~(\ref{dN_dErel}).}
    \label{fig:ebin}
\end{figure}

Once the system reaches equilibrium, the relative energy distribution is expected to follow the classical density of states, weighted by the Boltzmann factor to obtain the thermal distribution,
\begin{equation}
\frac{dN_{c \bar{c}}}{dE_{ c \bar{c}}} = (4\pi)^2 (2 \mu)^{\frac{3}{2}} C     \int_0^R dr r^2 \exp \left(-\frac{E_{c \bar{c}}}{T}\right) \sqrt{E_{c \bar{c}}-V(r)} \ ,    \label{dN_dErel}
\end{equation}
where $\mu=M_c/2$, $C$ is a normalization constant (fixed to the total number of pairs $N_{pair}$) and $R$ is the radius of a sphere with the same volume as the box in the simulation. Figure~\ref{fig:ebin} shows the resulting distribution of $E_{c \bar{c}}$ from the simulation compared to Eq.~(\ref{dN_dErel}) for $T=160$ MeV and volume $V=(8 \textrm{ fm})^3$. The simulations yield the correct equilibrium density of states. The left side of the distribution ($E_{\textrm{rel}}<0$) represents bound states. Accordingly, the right side corresponds to pairs with positive relative energy and therefore free charm and anticharm quarks. 

\begin{figure}[H]
    \centering
    \begin{subfigure}[b]{0.42\textwidth}  
        \centering
        \includegraphics[width=\linewidth]{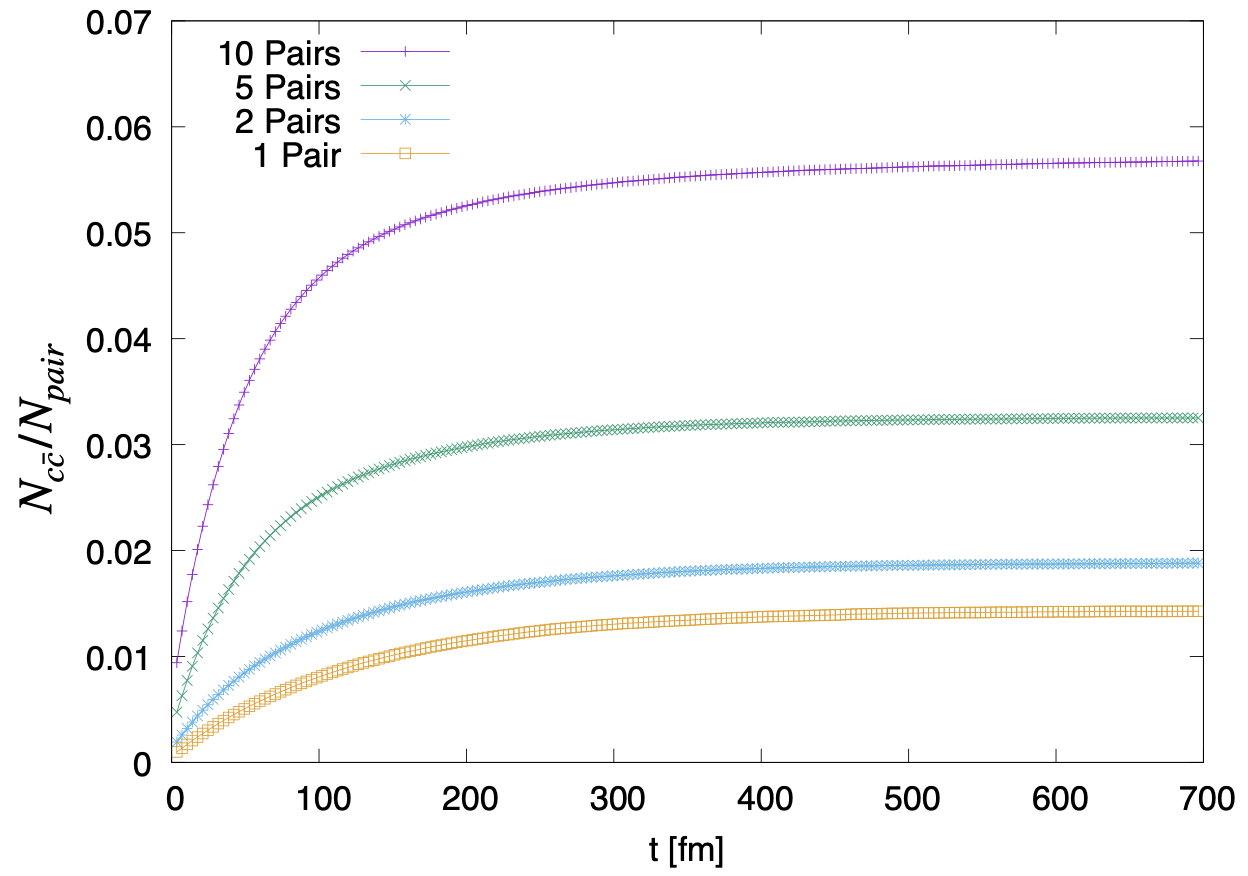}
        \caption{ }
        \label{fig:fractionBS}
    \end{subfigure}
    \hspace{0.05\textwidth}  
    \begin{subfigure}[b]{0.4\textwidth}
        \centering
        \includegraphics[width=\linewidth]{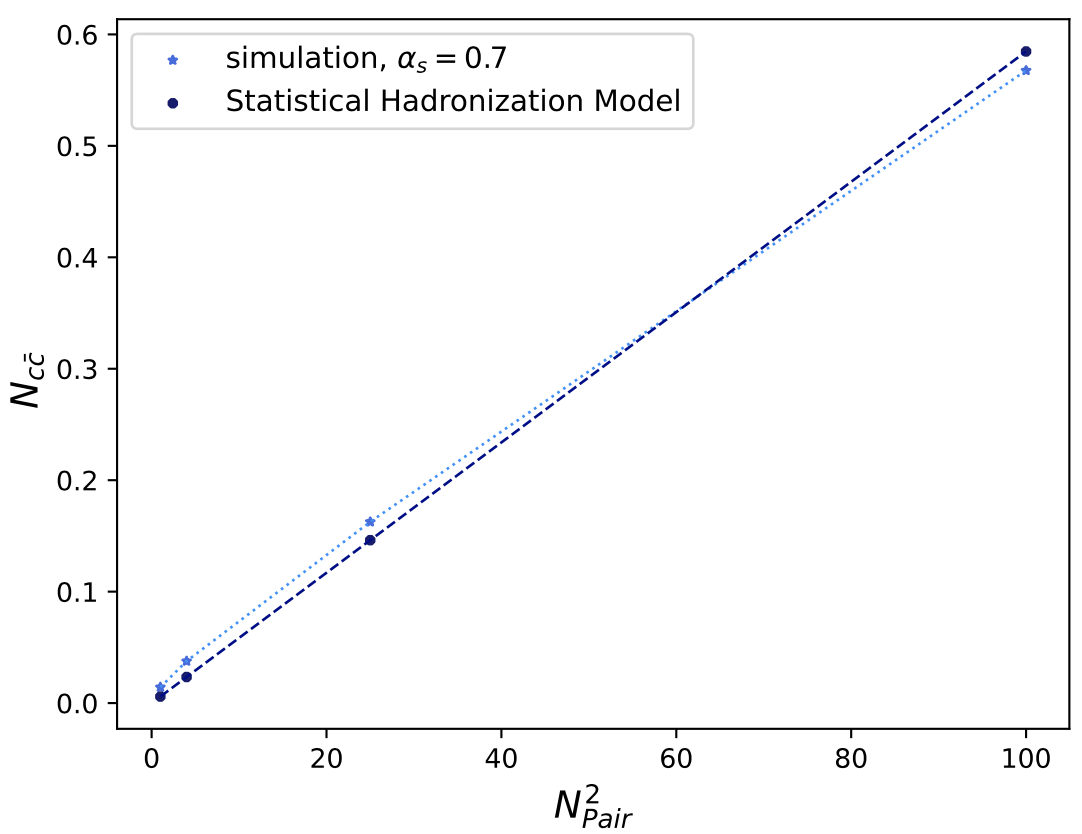}
       \caption{}
        \label{fig:SHM_comp}
    \end{subfigure}
    \caption{(a): Time evolution of the fraction of bound states for different numbers of initial HQ pairs in a box of volume $V=(10 \text{ fm})^3$ and temperature $T=160\ \text{MeV}$. (b): Final charmonium yields for different numbers of HQ pairs from the simulation compared to the results from the SHM.}
    \label{fig:BS-and-SHM}
\end{figure}

Furthermore, we study the time evolution of the bound-state formation for several numbers of initial pairs, as shown in the left part of Fig.~\ref{fig:fractionBS}.

The HQs are initially placed randomly in the box, which allows for a small chance of forming bound states immediately. Over time, recombination and dissociation processes occur until equilibration is reached, where their rates balance. Higher quark densities increase the probability of bound state formation, as shown by the rising equilibrium fraction with more pairs. To validate our results, we compare them to the SHM~\cite{Andronic:2003zv,Andronic:2017pug,Andronic:2021erx} in Fig.~\ref{fig:SHM_comp}, where we calculate the charmonium yield within the grand-canonical ensemble, according to
\begin{equation}
    N_{c \, \bar{c}} = \sum_{\alpha} \frac{N_c^2}{V} \frac{g_{\alpha}}{g_{c^2}}
    \left( \frac{2 \pi}{T} \right)^{3/2} \frac{M_{\alpha}^{3/2}}{M_c^3} \
    \exp \left( \frac{2M_c - m_{\alpha}}{T} \right) \quad ,  \qquad
      \alpha \in \{ J/\psi, \eta_c, \chi_c, \psi(2S) \} \ ,
\end{equation}
taking into account multiple charmonium states. To be consistent with the SHM results, we adjust the strong coupling constant, $\alpha_s(T_c)$, selecting a value of $\alpha_s = 0.7$, to align our results with the SHM predictions. With this choice the charmonium yield scales quadratically with the number of charm quark pairs.

To model realistic HICs we employ a dynamical description in an expanding background medium. A blast wave-like boost-invariant fireball is parametrized according to Ref.~\cite{VANHEES2015256}. 
The volume of the medium inside the fireball is given by
\begin{equation}
    V(\tau) = \pi a(\tau) b(\tau) (z_0 +c\tau)
\end{equation}
with the long and short semi-axes $a(\tau)$ and $b(\tau)$ expanding in a hyperbolic motion and $z_0 = c\tau_0$, where $\tau_0$ is the formation time of the fireball and $c=1$. The accelerations of the boundaries, $a_a$ and $a_b$, were chosen to achieve a good agreement with experimental data of $p_T$-spectra and elliptic flow of light hadrons~\cite{NaomiThesis}. In that way, different centrality classes can be parametrized. The flow field that quantifies the expansion is obtained from a combination of a transverse expansion from the velocities of the boundaries and the boost-invariant Bjorken flow for the longitudinal expansion~\cite{NaomiThesis}. 

The initial momentum distribution of the HQs in the fireball is given by PYTHIA, tuned to match the results from the Fixed-Order Next-to-Leading Logarithm calculations~\cite{Cacciari:1998it,Cacciari:2001td}, as done in Ref.~\cite{Song:2017phm}. The initial spatial distribution in the fireball volume follows from the Glauber model~\cite{NaomiThesis}. In this description, the elliptic flow of charm quarks and charmonium states is studied via the momentum components in the transverse plane of the collision ($p_x$, $p_y$),
\begin{equation}
v_2 = \bigg \langle \frac{p_x^2-p_y^2}{p_T^2}\bigg \rangle = \bigg \langle \frac{p_x^2-p_y^2}{p_x^2+p_y^2}\bigg \rangle \ .
\end{equation}

\begin{figure}[H]
    \centering
    \begin{subfigure}[b]{0.3\textwidth}  
        \centering
        \includegraphics[width=\linewidth]{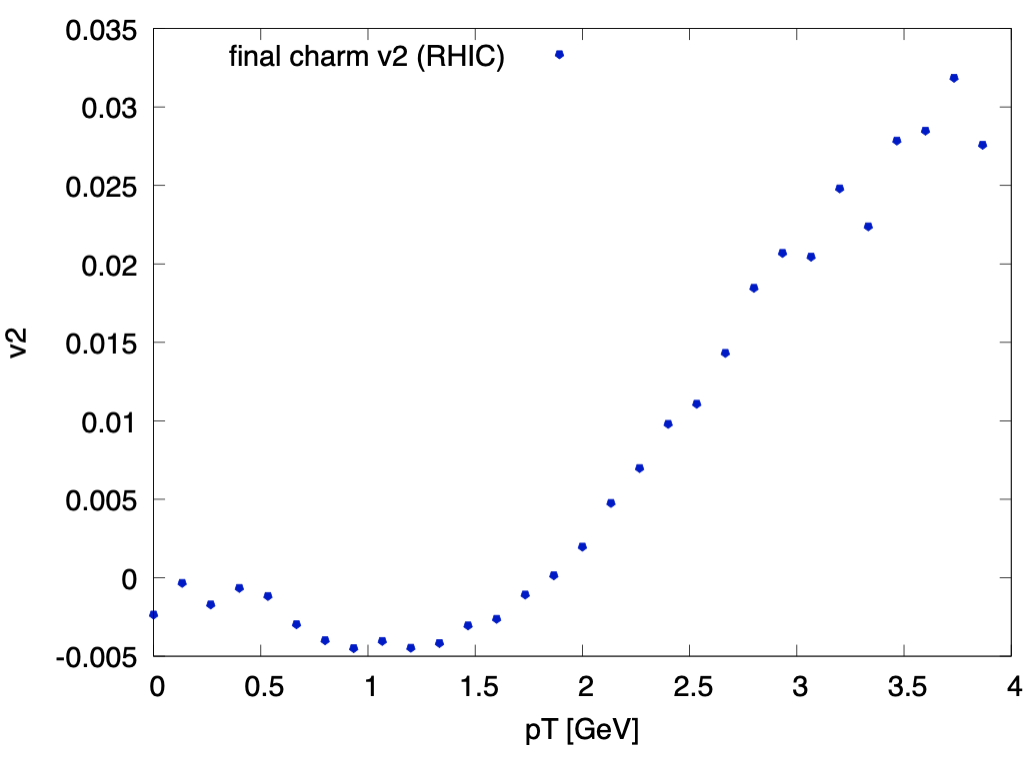}
        \caption{HQ $v_2$ for semi-central collisions ($20\%-40\%)$ at $\sqrt{s_{NN}}=200$ GeV (RHIC).}
        \label{fig:v2-charm-rhic}
    \end{subfigure}
    \hspace{0.03\textwidth}  
    \begin{subfigure}[b]{0.3\textwidth}
        \centering
        \includegraphics[width=\linewidth]{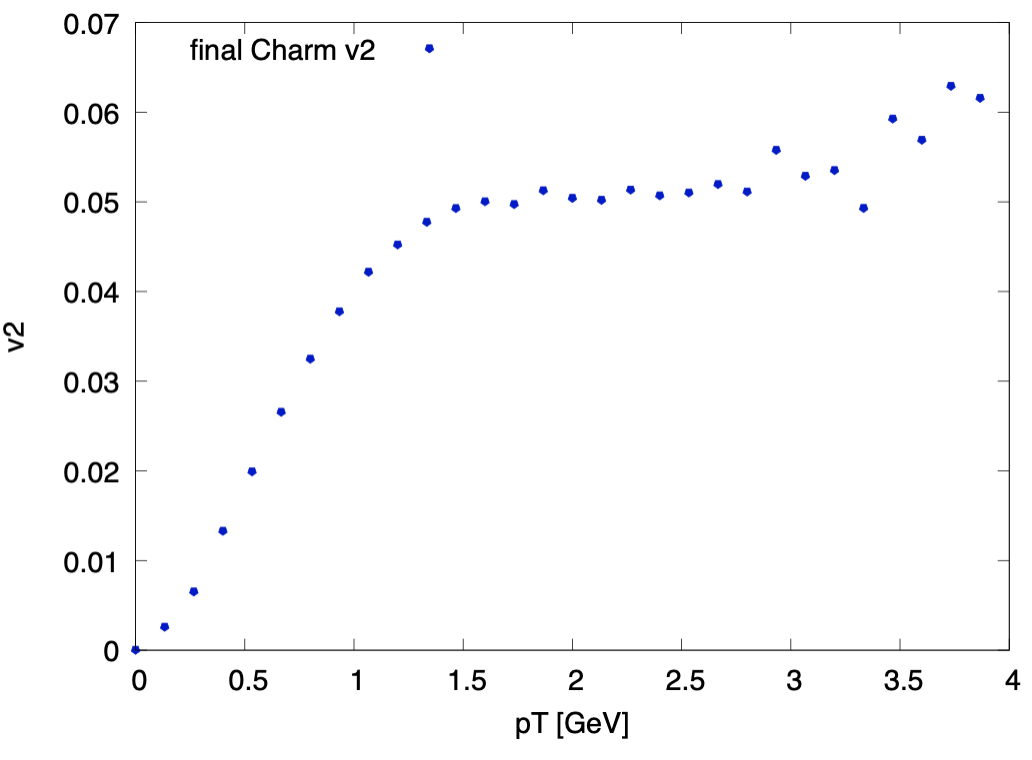}
        \caption{HQ $v_2$ for central collisions ($0\%-20\%)$ at $\sqrt{s_{NN}}=2.76$ TeV (LHC).}
        \label{fig:v2-lhc-charm1}
    \end{subfigure}
    \hspace{0.03\textwidth}
    \begin{subfigure}[b]{0.3\textwidth}
        \centering
        \includegraphics[width=\linewidth]{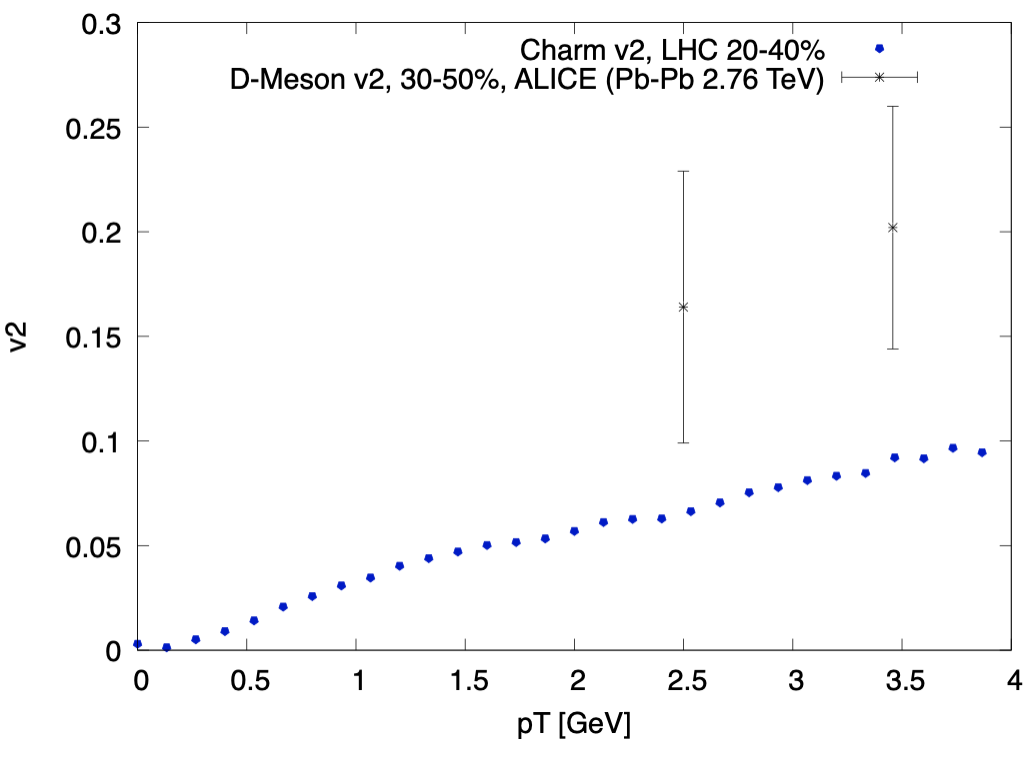}
        \caption{HQ $v_2$ for semi-central  collisions ($20\%-40\%)$ at $\sqrt{s_{NN}}=2.76$ TeV (LHC).}
        \label{fig:v2-lhc-charm2}
    \end{subfigure}
    \caption{Charm (anti-)quark elliptic flow as a function of $p_T$ for different collision energies and centralities.}
    \label{fig:v2-charm}
\end{figure}

Figure~\ref{fig:v2-charm} displays the elliptic flow of individual charm and anticharm quarks as a function of the transverse momentum for different energies and centralities. The simulations are carried out with $5$ initial HQ pairs and we consider fireball parametrizations for RHIC energies at $20-40\%$ centrality and for central ($0-20\%$) as well as semi-central ($20-40\%$) collisions at LHC energies. 

In all three cases $v_2$ is positive, indicating in-plane flow as charm and anticharm quarks are emitted preferentially along the reaction plane, with the flow increasing with transverse momentum. At $\sqrt{s_{NN}}=200$ GeV (left panel), a dip to negative values at low $p_T$ is observed, a typical feature for heavy particles, resulting from the interplay between transverse expansion and particle mass, where the HQs follow behind the bulk matter. A larger magnitude of $v_2$ is observed at higher energies since larger pressure gradients enhance the elliptic flow, with the highest values occurring in semi-central collisions. This highlights the sensitivity of $v_2$ to the initial spatial anisotropy. Our results serve as a proxy for open charm hadron flow, with comparisons to ALICE D-meson data at 30\%–50\% centrality~\cite{PhysRevLett.111.102301}. As expected, the charm quark $v_2$ is smaller, since D mesons receive additional flow contributions from hadronization and subsequent interactions.

\begin{figure}[H]
    \centering
    \begin{subfigure}[b]{0.3\textwidth}  
        \centering
        \includegraphics[width=\linewidth]{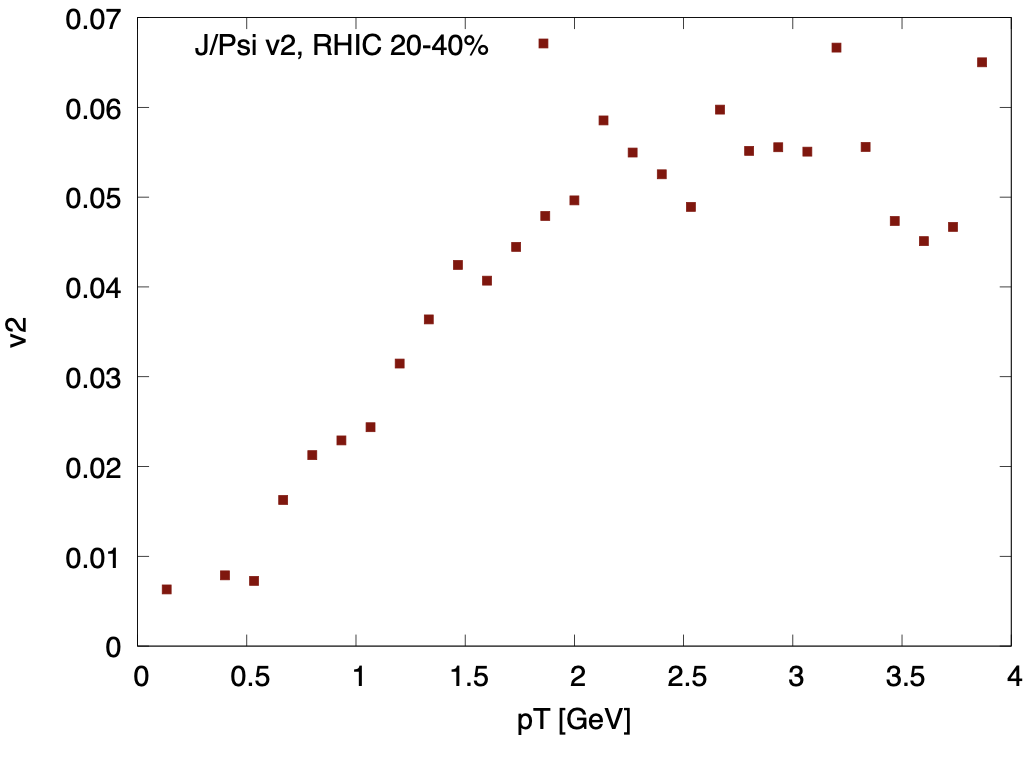}
        \caption{Semi-central collisions ($20\%-40\%)$ at $\sqrt{s_{NN}}=200$ GeV (RHIC).}
        \label{fig:v2-jpsi-rhic}
    \end{subfigure}
    \hspace{0.03\textwidth}  
    \begin{subfigure}[b]{0.3\textwidth}
        \centering
        \includegraphics[width=\linewidth]{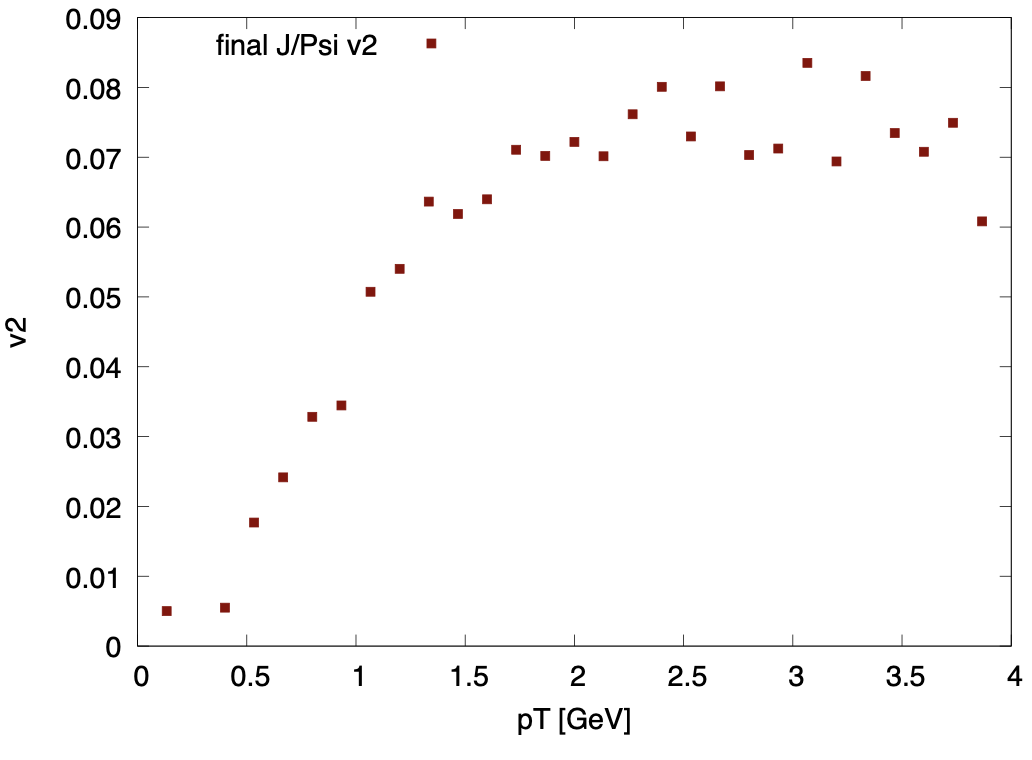}
        \caption{Central collisions ($0\%-20\%)$ at $\sqrt{s_{NN}}=2.76$ TeV (LHC).}
        \label{fig:v2-jpsi-lhc1}
    \end{subfigure}
    \hspace{0.03\textwidth}
    \begin{subfigure}[b]{0.3\textwidth}
        \centering
        \includegraphics[width=\linewidth]{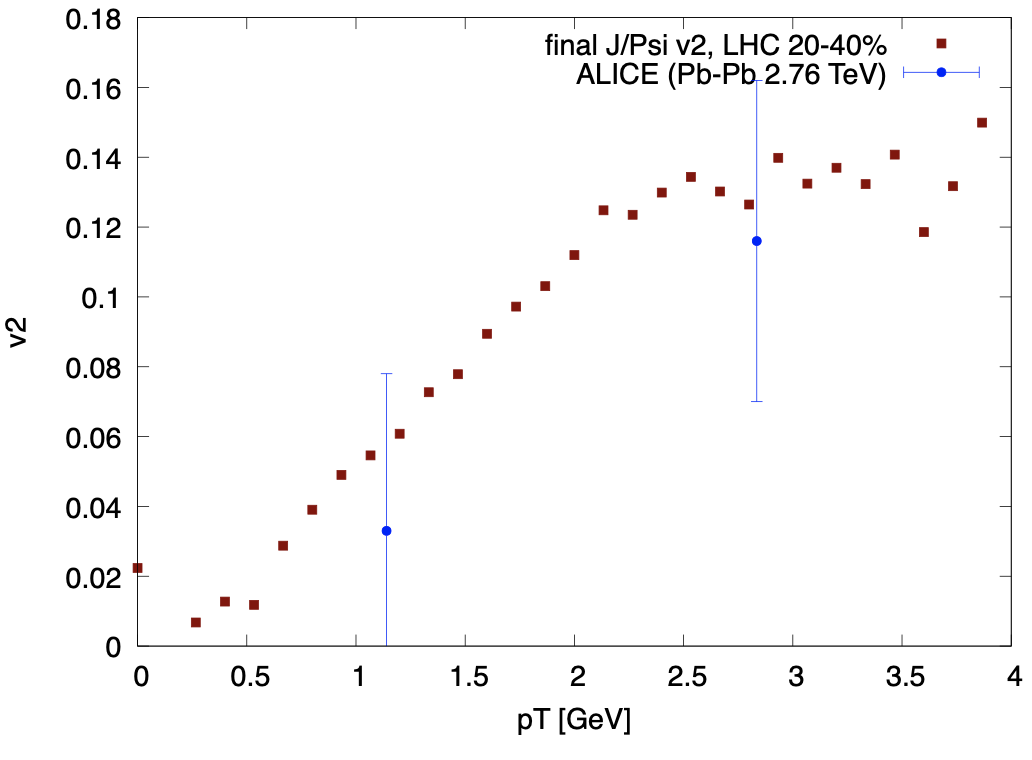}
        \caption{Semi-central collisions ($20\%-40\%)$ at $\sqrt{s_{NN}}=2.76$ TeV (LHC).}
        \label{fig:v2-jpsi-lhc2}
    \end{subfigure}
    \caption{Charmonium elliptic flow as a function of $p_T$ for different collision energies and centralities.}
    \label{fig:v2-charmonium}
\end{figure}

The elliptic flow of charmonia is investigated in Fig.~\ref{fig:v2-charmonium}. The results show a similar trend, with higher $v_2$ values at LHC energies, peaking in semi-central collisions. In all cases, the charmonium $v_2$ exceeds that of individual charm quarks. Since no primordial charmonium is included in the simulation, only the flow resulting from regenerated HQs is considered. The regenerated charmonia inherit the HQ flow from the recombination process, leading to an overall higher magnitude. Although improved statistics are needed, the results align with experimental data. For example, the comparison with Pb+Pb collisions at $\sqrt{s_{NN}} =2.76$ TeV from the ALICE experiment at forward rapidity~\cite{PhysRevLett.111.162301,hepdata.61768} shows that the $v_2$ of inclusive $J/\psi$ reaches up to $11.6\%$ for 20–40\% centrality, matching well within uncertainties.

\section{Conclusions and Outlook}
\label{sec:outlook}
We developed a microscopic model to describe charmonia based on relativistic Langevin simulations, which we tested in box simulations at fixed temperature and ensured that the correct equilibrium distribution is reached. We observed that bound-state formation, dissociation, and regeneration occur in the expected manner by comparing the final charmonium multiplicities to the SHM. Subsequently, we embedded our framework in an expanding medium, modeled by an elliptic fireball. Within this dynamical description we investigated the elliptic flow of charm quarks and charmonia and obtained a reasonable comparison to ALICE data. Future improvements include the calculation of the nuclear modification factor and the incorporation of primordial charmonia, which will allow for a more accurate comparison to experimental data. Finally, the description can be generalized to consider bottom flavor to cover the bottomonium sector.

\acknowledgments

We thank Taesoo Song for helpful assistance in implementing the PYTHIA initial distribution. We acknowledge support by the Deutsche Forschungsgemeinschaft (DFG) through the CRC-TR 211 “Strong- interaction matter under extreme conditions”. J.T.-R. acknowledges funding from project numbers CEX2019-000918-M (Unidad de Excelencia “María de Maeztu”), PID2020-118758GB-I00 and PID2023-147112NB-C21, financed by
the Spanish MCIN/ AEI/10.13039/501100011033/.

\bibliographystyle{JHEP}
\bibliography{references}

\end{document}